\documentclass[%
reprint,
 amsmath,amssymb,
 aps,
prb,
]{revtex4-2}

\usepackage{graphicx}% Include figure files
\usepackage{dcolumn}% Align table columns on decimal point
\usepackage{bm}% bold math
\usepackage{amsmath,empheq}
\usepackage{graphicx}
\usepackage[colorlinks=true, allcolors=blue]{hyperref}
\usepackage{braket}
\usepackage{soul, xcolor}
\sethlcolor{yellow}

\newcommand{\nnpp}{nextnano\raisebox{0.1ex}{\footnotesize ++}}

\begin{document}

\title{Magneto-optical spectroscopy of (In)GaSb/AlGaSb quantum dots}

\author{
G. Barbieri$^{1, 2}$,
J. M. Llorens$^{1}$,
A. Chellu$^{3}$,
J. Hilska$^{3}$,
L. Leguay$^{4}$,
A. Schliwa$^{4}$,
T. Hakkarainen$^{3}$,
and B. Al\'en$^{1}$\\[0.5em]
{\small\itshape
$^{1}$Instituto de Micro y Nanotecnolog\'ia, IMN-CNM, CSIC, Calle de Isaac Newton 8, E-28760 Tres Cantos, Spain\\
$^{2}$Departamento de Física de Materiales, Facultad de Ciencias Físicas, Universidad Complutense de Madrid, E-28040 Madrid, Spain\\
$^{3}$Optoelectronics Research Centre, Physics Unit, Korkeakoulunkatu 3, 33720 Tampere, Finland \\
$^{4}$Institut f\"ur Festk\"orperphysik, Fakult\"at II, Technische Universit\"at Berlin, Hardenbergstra{\ss}e 36, 10623 Berlin, Germany
}
}

\begin{abstract}
(In)GaSb/AlGaSb quantum dots combine high uniformity and low strain with single-photon telecom emission, but understanding the full electronic structure remains essential for controlling the optical properties beyond the emission wavelength. Here, we combine circularly resolved magneto-photoluminescence with three-dimensional eight-band $k\cdot p$ calculations to investigate the orbital structure underlying their excited-state emission. The ground- and excited-state energies constrain an effective confinement geometry whose calculated optical transitions reproduce the principal measured magnetic dispersions, following distinct electron orbital families with secondary contributions from closely spaced and mixed hole states. Calculations further show that increasing In content makes the excited hole levels more widely spaced and predominantly heavy-hole in character, but may also draw the hole wavefunction out of the dot and toward the surrounding quantum well. These findings establish an experimentally constrained description of the electronic structure of (In)GaSb/AlGaSb quantum dots and highlight the importance of considering valence-band confinement when engineering the optical and magnetic properties of telecom quantum emitters.
\end{abstract}

\maketitle

\section{Introduction} \label{sec:intro}

% Motivation and state of (In)GaSb/AlGaSb LDE
Semiconductor quantum dots (QDs) provide deterministic sources of single and entangled photons, while direct emission in the telecom bands is particularly relevant for fiber-based quantum networks and photonic integration \cite{yu2023telecom}. Local droplet etching (LDE) followed by nanohole infilling offers a complementary route to conventional strain-driven self-assembly, enabling low-density, low-strain, highly symmetric emitters whose quantum-optical performance is well established for the GaAs/AlGaAs system \cite{huber2017highly, gurioli2019droplet, da2021gaas}. Extending this approach to GaSb-based heterostructures combines these advantages with direct telecom-wavelength emission \cite{chellu2021highly} and with the broader potential of antimonide materials for monolithic integration on silicon \cite{huang2008simultaneous, rodriguez2016x, tournie2022mid}. single QD spectroscopy has established single photon emission from strain-free GaSb/AlGaSb QDs in the telecom S-band \cite{michl2023strain}, while more recent work has linked nanohole morphology and carrier-transfer dynamics directly to their single-photon performance \cite{masson2026engineering}.

% Predictive understanding of the confined electronic structure
In GaAs/AlGaAs LDE QDs, combining structural characterization with multiband modelling has enabled quantitative reproduction of optical transitions \cite{graf2014excitonic} and predictive design of electric and magnetic properties \cite{schimpf2025optical}. Comparable progress is being made in nanohole-filled (In)GaSb/AlGaSb QDs, where varying the In content and the nominal thickness of (In)GaSb deposited during nanohole infilling has enabled GS emission tuning from approximately $1480$ to $1550~\mathrm{nm}$, while maintaining ensemble inhomogeneous broadening below $7~\mathrm{meV}$. Structurally detailed $k\cdot p$ and configuration-interaction calculations accurately reproduce the emission shifts associated with both growth parameters \cite{peil2026predictive}, but understanding the excited-state structure is nevertheless essential for extending this control beyond the GS emission wavelength. Previous modelling of GaSb/AlGaSb QDs has already demonstrated the sensitivity of optical transitions and the competition between $\Gamma$- and L-valley electron states to confinement geometry \cite{leguay2024unveiling}. Moreover, quasi-resonant single QD spectroscopy and multiband calculations revealing closely spaced hole states, possibly with substantial heavy-hole-light-hole mixing, contributing to excited-state absorption \cite{hakkarainen2026telecom}, consistent with the weak strain and comparatively large dimensions of the dots. Previous modelling of GaSb/AlGaSb QDs has demonstrated the sensitivity of optical transitions and the competition between $\Gamma$- and L-valley electron states to confinement geometry \cite{leguay2024unveiling}. Quasi-resonant single QD spectroscopy, interpreted with multiband calculations, has further identified contributions from closely spaced hole states to excited-state absorption \cite{hakkarainen2026telecom}. Magneto-photoluminescence provides an additional probe of this electronic structure because the magnetic-field dependence of optical transitions is sensitive to the orbital character and spatial extent of the confined states. Magneto-optical studies of In(Ga)As/GaAs QDs have resolved excited-state shell structure and revealed its angular-momentum content \cite{raymond2004excitonic, awirothananon2008single}, while measurements and calculations of GS and excited-state diamagnetic shifts in unstrained GaAs/AlGaAs QDs have established a quantitative connection between magnetic response and confinement geometry \cite{sidor2006high}.

% Scope and contribution of this work
Here, we combine polarization resolved magneto-photoluminescence with three-dimensional eight-band $k\cdot p$ calculations to investigate the electronic structure of (In)GaSb/AlGaSb QDs. We use the zero-field GS energy and excited-state splittings to constrain an effective confinement geometry, then test the resulting model against the measured magnetic-field dependence of the optical transitions. Across all three samples, the calculations reproduce the principal experimental dispersions, which primarily reflect distinct electron orbital families, with additional contributions from closely spaced and mixed hole states. We further examine how In incorporation modifies the energy separations, valence-band character, and spatial localization of the hole states. These results connect the observed optical transitions to the underlying carrier confinement, providing a basis for engineering optical properties beyond the GS emission wavelength.

% Connected QD-QW system
%The nanohole-filling geometry does not produce isolated QDs: the deposited GaSb or InGaSb also forms a thin quantum well (QW) between the nanoholes, resulting in a connected QD-QW heterostructure.  Carrier capture from such a reservoir depends on its energetic structure and density of states \cite{syperek2013impact}, while thermal redistribution and magnetic-field-dependent transport can further modify the populations delivered to the QDs \cite{lee1997temperature, larsson2006magnetic}. Time-resolved measurements on GaSb/AlGaSb QDs have revealed slow QD population following above-barrier excitation, attributed to carrier transfer through the surrounding QW \cite{michl2023strain}, while temperature-dependent spectroscopy and rate-equation modelling have identified thermally activated capture and escape processes between the QW and QD states \cite{leguay2024unveiling}. More recently, Masson \textit{et al.} directly linked the nanohole morphology to QD-QW charge-transfer dynamics and showed that the resulting carrier capture strongly affects the single QD optical response \cite{masson2026engineering}. In our companion study of the present samples, we further find that carriers relax through disorder-localized QW states and acquire a magnetic-field-dependent spin imbalance before capture by the QDs \cite{barbieri2026magneto}.
\section{Samples and Methods} \label{sec:methods}

\begin{figure}[h!]
    \centering
    \includegraphics[width=0.99\linewidth]{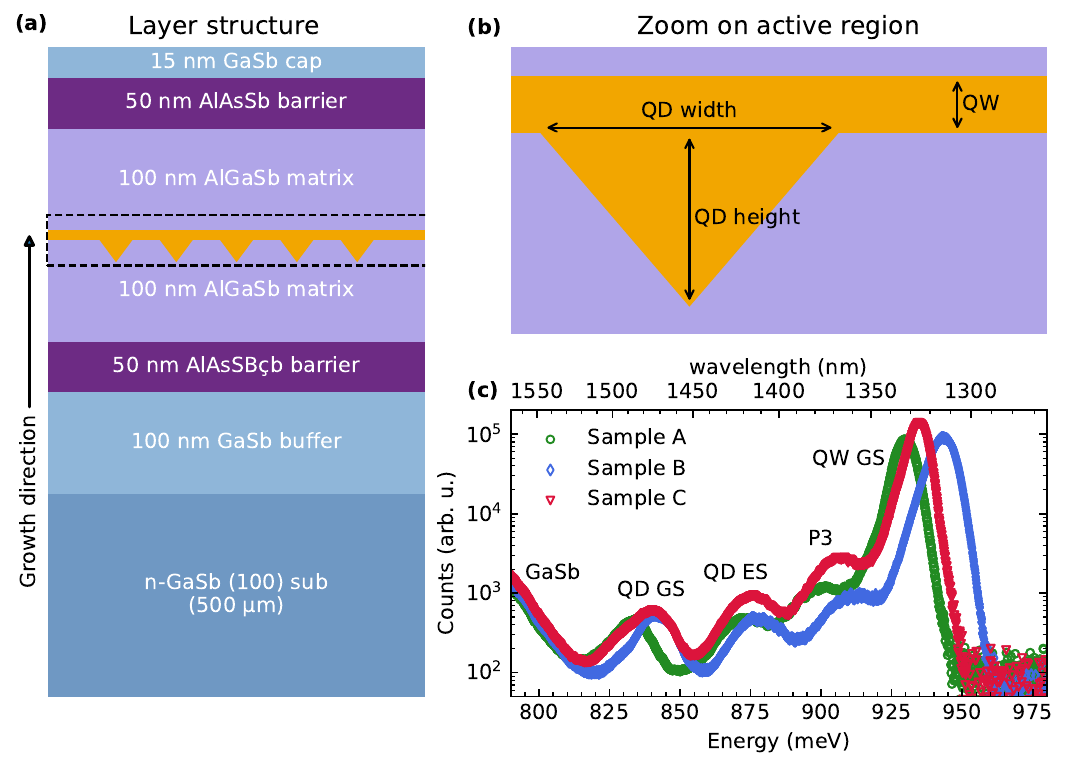}
    \caption{
    Overview of the investigated (In)GaSb/AlGaSb QD structures.
    (a) Layer structure of the samples.
    (b) Active-region geometry defining the QD width, QD height, and the thickness of the overlying QW used in the calculations.
    (c) Zero-field PL spectra of Samples A-C at $4~\mathrm{K}$, showing the QD ground-state (QD GS), QD excited-state (QD ES), P3, and QW ground-state (QW GS) emission.
    }
    \label{fig:overview}
\end{figure}

% Growth and AFM
The investigated heterostructures were grown by solid-source molecular beam epitaxy on $n$-GaSb(100) substrates. The layer structure follows the GaSb/AlGaSb LDE platform outlined in \cite{chellu2021highly}: a $100~\mathrm{nm}$ GaSb buffer is followed by a $50~\mathrm{nm}$ lattice-matched AlAsSb lower cladding layer and a $100~\mathrm{nm}$ AlGaSb matrix, in which nanoholes are formed by Al-droplet etching. After nanohole formation, the structures investigated here were refilled with GaSb or InGaSb; the deposited refill material also covers the planar surface between the nanoholes, forming the surrounding quantum well (QW) shown in Figure~\ref{fig:overview}(a,b). The active region is capped by a further $100~\mathrm{nm}$ AlGaSb layer, a $50~\mathrm{nm}$ AlAsSb upper cladding layer, and a $15~\mathrm{nm}$ GaSb cap. Atomic force microscopy characterization of the uncapped nanoholes yielded an areal density of approximately $2.6\times10^{7}~\mathrm{cm}^{-2}$, an average depth of about $12~\mathrm{nm}$, and an average width of about $50~\mathrm{nm}$ \cite{leguay2024unveiling}.  Samples~A, B, and C were grown with nominal (In)GaSb deposition thicknesses of $16.0$, $18.6$, and $20.0$ monolayers, respectively, with $3\%$ In incorporated only in Sample~A, as summarized in Table~\ref{tab:structural_parameters}.

\begin{table}[h!]
\caption{\label{tab:structural_parameters} Structural parameters of the investigated (In)GaSb/AlGaSb QD samples. The thickness denotes the nominal equivalent thickness of (In)GaSb deposited during nanohole infilling, expressed in monolayers (ML) and nanometres.}
\begin{ruledtabular}
\begin{tabular}{cccc}
Sample & Thickness & Active Region & Barrier Al content \\
& (ML / nm) & & (\%) \\
\hline
A & 16.0 / 4.88 & $\mathrm{In}_{0.03}\mathrm{Ga}_{0.97}\mathrm{Sb}$ & 32.5 \\
B & 18.6 / 5.67 & $\mathrm{GaSb}$ & 31.9 \\
C & 20.0 / 6.10 & $\mathrm{GaSb}$ & 31.3 \\
\end{tabular}
\end{ruledtabular}
\end{table}

% MPL
The magneto-photoluminescence dataset analysed here is also examined in our companion study, which investigates disorder-mediated carrier relaxation through the quantum well under applied magnetic fields \cite{barbieri2026magneto}. Here, we focus on the lower-energy QD ground- and excited-state (GS and ES) emission. Measurements were performed at 4 K with magnetic fields up to 9 T applied along the growth direction. The samples were excited non-resonantly with a linearly polarized $785$-nm continuous-wave laser. The emission was collected in back-scattering configuration and dispersed by a $300$-mm-focal-length monochromator before detection with a Peltier-cooled InGaAs photomultiplier. Representative zero-field spectra of Samples A-C are shown in Figure~\ref{fig:overview}(c), comprising emission from the GaSb substrate, the QD GS and ES, the P3 feature tentatively associated with localized states near the nanohole rim \cite{ulhaq2016vanishing}, and the QW GS. The $\sigma^{+}$ and $\sigma^{-}$ emission components were obtained by selecting a fixed circular polarization in detection and reversing the magnetic-field direction while keeping the optical path unchanged.

% Simulations
We calculated the electronic structure using the commercially available three-dimensional eight-band $k\cdot p$ model \nnpp \cite{birner2007nextnano, nextnano}. Strain and piezoelectric fields were included in the calculations, including second-order piezoelectric contributions. The strain calculation used periodic boundary conditions in the growth plane, with the substrate constraint along the growth direction imposed by the GaSb lattice parameter. Material parameters were taken from the nextnano++ database. Following the approach of Leguay et al. for nanohole-filled GaSb/AlGaSb QDs \cite{leguay2024unveiling}, the QW and QDs were constructed with the simplified inverted-conical geometry illustrated in Figure~\ref{fig:overview}(b), treating the QD height and width as effective confinement parameters. During the geometry and composition sweeps, the overlying QW thickness was fixed at $6~\mathrm{nm}$ and the Al content of the AlGaSb matrix at $32\%$. For the sample-specific calculations, the nanohole measurements were kept fixed while the QW thickness was set to the nominal thickness of each sample, with the AlGaSb matrix composition was set at $32\%$. Sample~A was modelled with it's nominal In content of $3\%$. 

For each structure, we calculated the confined electron and hole states, the interband transition energies, and the circularly polarized optical matrix elements. We quantify the transition strength by the transition intensity $I_{\sigma^\pm}$, defined as the squared electron-hole optical matrix element for $\sigma^\pm$ polarization (see \nnpp documentation~\cite{nextnano} for details), which is related to the oscillator strength by $f_{\sigma^\pm}=I_{\sigma^\pm}/E$, where $E$ is the transition energy. Single-particle energies are referenced to the HH band-edge of the AlGaSb barrier. The envelope orbital angular momentum along the growth direction was evaluated separately for each Bloch component of the multiband wavefunction using $\hat{L}_z=-i\hbar\,\partial/\partial\phi$, where $\phi$ is the azimuthal coordinate around the axis of the inverted cone. We denote the corresponding orbital angular-momentum projection in units of $\hbar$ by $l_z$, and classify the electron states according to their dominant $l_z$ character~\cite{van2011g}. The calculated transition energies are shifted downward by a sample-dependent, field-independent offset of $8$--$12~\mathrm{meV}$ to match the experimental zero-field GS emission. The magnitude of this offset is consistent with the excitonic corrections obtained from configuration-interaction calculations for similar (In)GaSb/AlGaSb QDs~\cite{peil2026predictive}.
\section{Results} \label{sec:results}

\subsection{Magneto-photoluminescence spectra}
\label{sec:results_magnetopl}

\begin{figure}%[h!]
    \centering
    \includegraphics[width=0.99\linewidth]{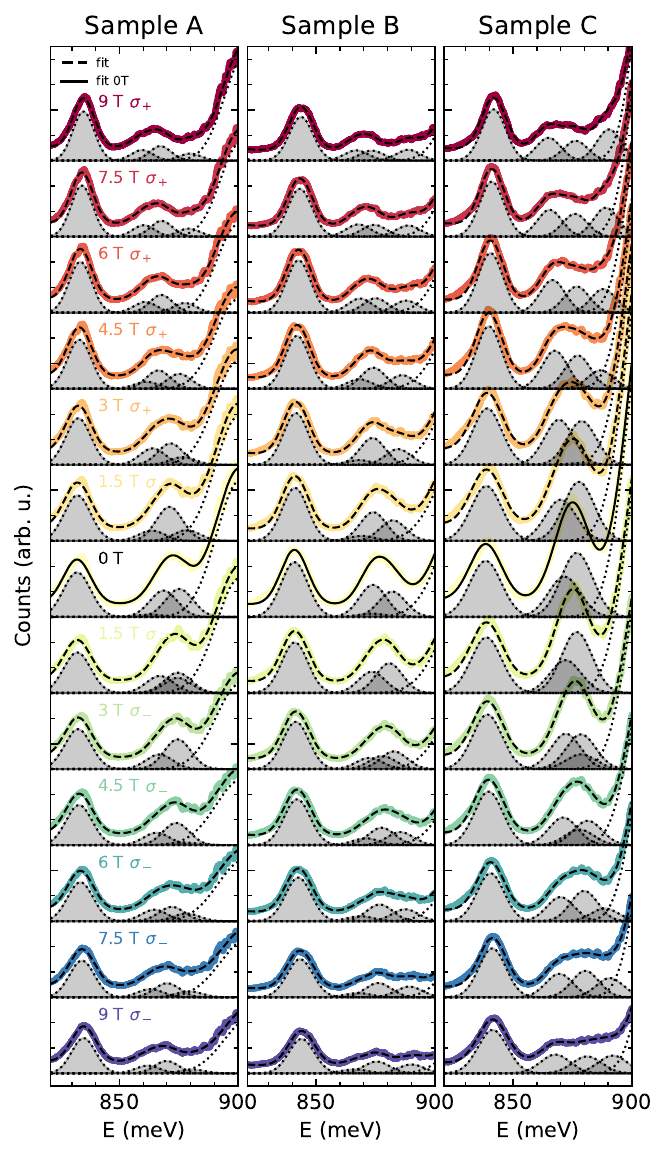}
    \caption{Magnified view of the QD MPL spectra of Samples A-C, vertically offset and ordered, top to bottom, from $9$~T $\sigma^{+}$ to $9$~T $\sigma^{-}$. At each magnetic field, the QD emission is decomposed into one GS and three ES Gaussian components, all constrained to a common linewidth. Colored curves show the measured spectra, dashed black curves (solid for 0 T) the total fits, and dotted gray curves the individual fitted components. Additional lineshapes account for the tails of the GaSb bulk and P3 emission, while a constant background term approximates the broad underlying spectral background.}
    \label{fig:magnetoPL}
\end{figure}

% MPL description
Figure~\ref{fig:magnetoPL} shows a magnified view of the QD spectral region as a function of magnetic field applied along the growth direction for all samples. At zero magnetic field, the QD GS bands of Samples A, B, and C are centered at 832, 841, and 838~meV, respectively, while the corresponding ES envelopes are centered approximately 39, 36, and 36~meV above the GS. In all three samples, the ES emission is broader than the GS and shows no clearly resolved internal structure at zero field. With increasing magnetic field $B$, the GS remains a single inhomogeneously broadened band, whereas the ES line shape becomes progressively more structured. Shoulders and, at some fields, weak local maxima emerge within the broad ES envelope, with similar overall behavior observed across Samples A--C.

% Deconvoltion
To quantify this evolution, each spectrum was fitted with one Gaussian component for the QD GS and three for the ES manifold, as illustrated in the magnified QD spectral region shown in Figure~\ref{fig:magnetoPL}. Two ES components were insufficient to reproduce the field-dependent evolution, particularly at higher fields, whereas three provided a consistent description across the measured series. Across the wider fitting window, two additional lineshapes accounted for the GaSb emission tail at low energies and the P3 emission tail at high energies. At each magnetic field, the four QD components shared a common linewidth, while their fitted energies and amplitudes varied independently. A constant background accounted phenomenologically for the broad underlying emission across the fitting window \cite{chellu2021highly}.

% Deconvoltion result description TO BE MOVED IN DISCUSSION
% Figure~\ref{fig:experiment_theory} (a-c) repeats the experimental QD spectra of Samples~A-C at selected magnetic fields, overlaid with electron-hole pairs from simluations, as detailed in the next section. In Figure~\ref{fig:experiment_theory} (d-f), the extracted experimental Gaussian energies and widths are compared to simlations. The shaded bands are centred at the fitted energies, with their thickness given by $\mathrm{FWHM}\times A_i/A_{\mathrm{GS}}$, reflecting the amplitude of each component relative to the GS at the same field. At low magnetic fields, the three ES contributions remain largely unresolved within the broad emission envelope, but their extracted energies follow increasingly distinct trajectories as $B$ increases. One exhibits a comparatively weak dispersion, while the other two separate more strongly, with the lowest and highest ES energies differing by approximately $25~\mathrm{meV}$ at $B=9$~T. At higher fields, the high-energy ES contribution becomes less reliably determined because of increasing overlap with the low-energy tail of the P3 feature.

\subsection{Electronic structure calculation}
\label{sec:results_calculations}

\begin{figure}%[h!]
    \centering
    \includegraphics[width=0.99\linewidth]{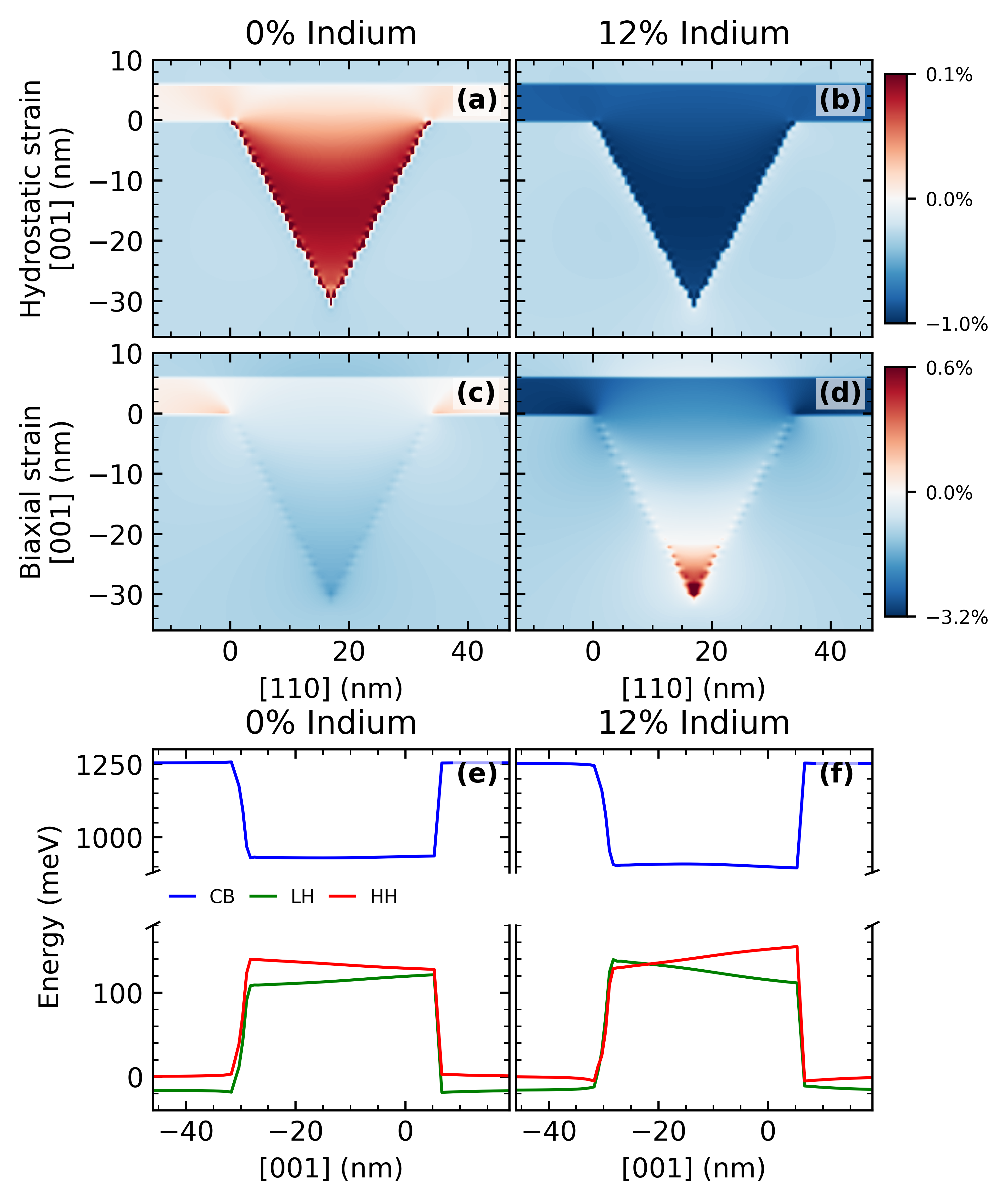}
    \caption{
    Calculated strain and band-edge profiles for a QD of width $34~\mathrm{nm}$ and height $31~\mathrm{nm}$ embedded in a $6~\mathrm{nm}$ QW. The In fraction in the (In)GaSb QD and QW is $0\%$ (left) or $12\%$ (right). Panels (a, b) and (c, d) show the hydrostatic and biaxial strain, respectively, in a [110]–[001] cross section. Panels (e, f) show the conduction-band (CB), heavy-hole (HH), and light-hole (LH) bandedges along [001] through the QD centre.
    }
    \label{fig:Sims_input}
\end{figure}

\begin{figure}%[h!]
    \centering
    \includegraphics[width=0.99\linewidth]{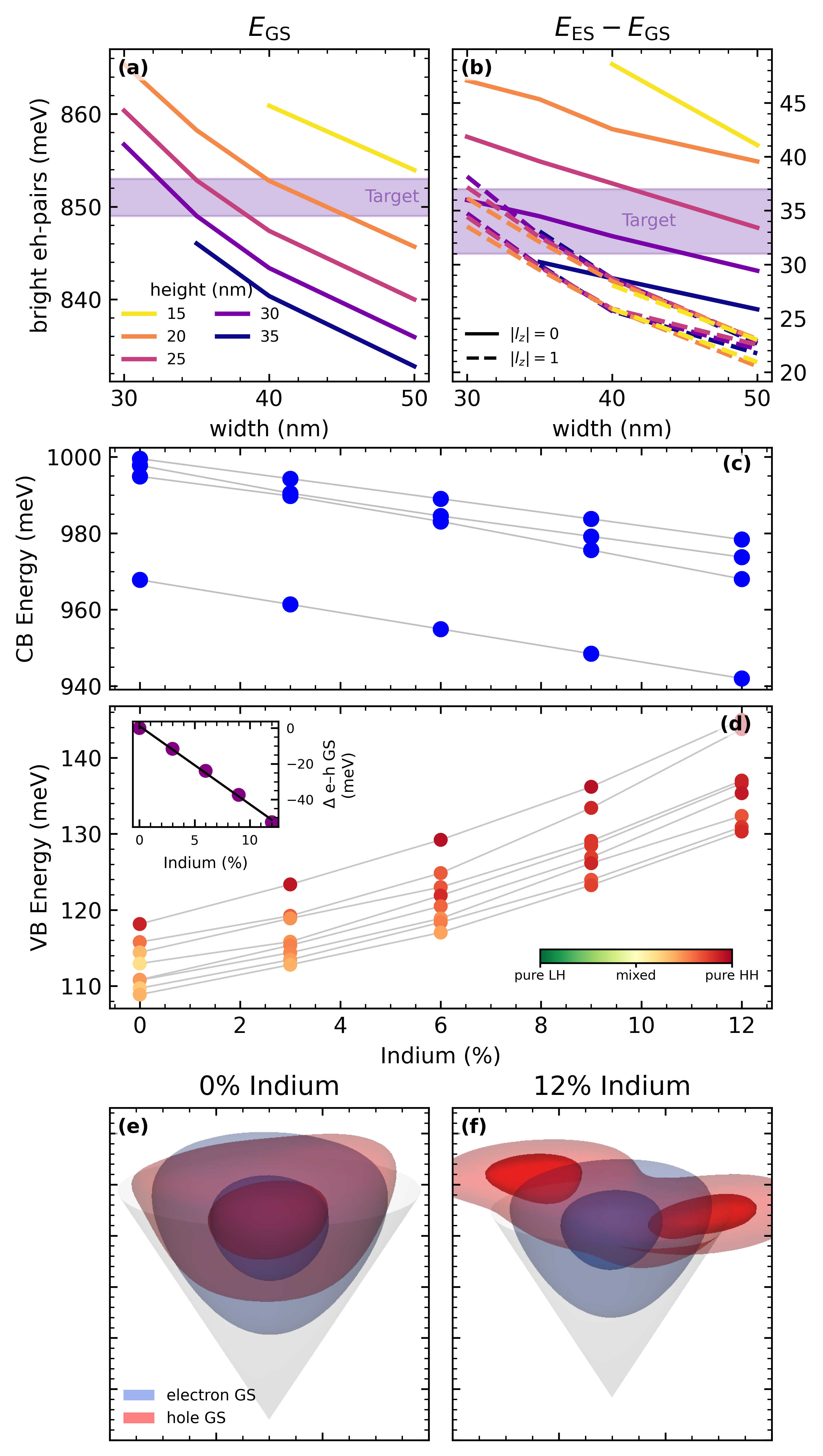}
    \caption{
    Calculated confined states and optical transitions of (In)GaSb QDs embedded in a $6~\mathrm{nm}$ QW. For In-free QDs, (a) shows the GS transition energy and (b) the separation of the brightest excited-state transitions ES from the GS as a function of QD width. Colours denote QD heights from $15$ to $35~\mathrm{nm}$. In (b), dashed and solid lines identify transitions involving $|l_z|\approx1$ and $l_z\approx0$ electron states, respectively. Shading marks the Sample~C target ranges; the GS target in (a) includes an upward rigid shift of $8$--$12~\mathrm{meV}$ applied to the measured emission energy (see Section~\ref{sec:methods}). For a QD of $34~\mathrm{nm}$ width and $31~\mathrm{nm}$ height, (c,d) show the electron and hole energies as the In content of the QD and QW varies together. Colour in (d) indicates the HH--LH composition, and the inset shows the GS transition energy relative to the In-free case. Panels (e,f) show the electron (blue) and hole (red) GS probability densities at $0\%$ and $12\%$ In, respectively; the isosurfaces enclose $10\%$ and $90\%$ of each density, and the grey shading indicates the QD geometry.
    }
    \label{fig:Sims_output}
    
\end{figure}

% GaSb: Strain and bandedges
The GaSb QD is lattice matched to the GaSb substrate and has a mismatch of $0.24\%$ with the surrounding AlGaSb matrix, significantly smaller than in other systems such as InAs/GaAs ($-6.7\%$). Alloying the dot with $12\%$ In, i.e. In$_{0.12}$Ga$_{0.88}$Sb, increases the mismatch to $-0.52\%$ and changes its sign. Figs.~\ref{fig:Sims_input}(a,b) show cross-sections of the hydrostatic strain, $e_\mathrm{h}=e_{xx}+e_{yy}+e_{zz}$, for 0\% and 12\% In, respectively. Its average value in the QD volume changes from $<0.1\%$ (tensile) to $-1.0$\% (compressive). The biaxial strain, $e_{\mathrm{biax}}=e_{xx}+e_{yy}-2e_{zz}$, has opposite gradients along the axis of the two QDs (Figure~\ref{fig:Sims_input}(c,d)). With GaSb it is negative ($-1.6\%$) at the nanohole tip and approaches zero towards the QW. At $12\%$ In it reaches approximately $+0.6\%$ at the tip, changes sign within the dot, and becomes increasingly negative towards the QW, with values up to $-3.3\%$ whithin the QW regions. The CB edge shifts downward by roughly $30~\mathrm{meV}$ on average and remains nearly flat across the filling (Figure~\ref{fig:Sims_input}(e,f)). The HH edge rises by about $10~\mathrm{meV}$ on average, but the reversal of the biaxial strain gradient also shifts its maximum from the nanohole tip to the QW. This maximum rises from approximately $140$ to $155~\mathrm{meV}$, moving the minimum of the hole confinement potential towards the dot opening.

%Figure~\ref{fig:Sims_input}(c-d) the biaxial component, that will affect the heavy-hole and light-hole bands splitting. The calculated hydrostatic strain then reaches $+0.4\%$ in the QD (Figure~\ref{fig:Sims_input}(a)). %For comparison, self-assembled InAs/GaAs and telecom-emitting InAs/InP QDs have lattice mismatches of $6.6\%$ and $3.1\%$, respectively \cite{cornet2006electronic}, with calculations for InAs lens-shaped QDs in the two matrices giving hydrostatic strains of approximately $-7\%$ and $-3\%$ near the QD centre \cite{gong2008electronic}.
%Here, the residual biaxial strain reaches its minimum of $-1.9\%$ at the nanohole tip. Thus, the HH edge is highest near the tip, favouring hole confinement in this region, while the separation from the LH edge decreases moving towards the nanohole base (Figure~\ref{fig:Sims_input}(c,e)).

% GaSb: Energies
After diagonalizing the Hamiltonian, we obtain the electron and hole states. Fig.~\ref{fig:Sims_output}(a) shows the energy of the lowest-energy (GS) electron-hole transition as a function of QD width for different heights, and Fig.~\ref{fig:Sims_output}(b) the energies of the higher transitions (ES) relative to the GS. In these relatively large QDs, the hole states are closely spaced owing to their heavy effective mass, so the spacings between transitions are determined mainly by the electron states. We therefore label the ES transitions by the dominant envelope angular momentum $l_z$ of the electron state, as defined in Section~\ref{sec:methods}. Increasing either the width or the height of the GaSb QD redshifts the GS transition, which involves an electron state with $l_z\approx0$ in all cases, because of the weaker confinement. For the ES, we distinguish transitions involving $|l_z|\approx1$ states (dashed lines) from those with $l_z\approx0$ (solid lines). The $|l_z|\approx1$ states are $p$-like and depend more strongly on the QD width, whereas the $l_z\approx0$ ES have an additional node along the $z$ axis, i.e. they are vertically excited states, and are therefore more sensitive to the QD height. 

% Sweetspot
%To constrain the effective QD dimensions, we used the zero-field GS emission energy and ES-GS separation of Sample~C, whose nominal filling thickness is closest to the $6~\mathrm{nm}$ QW thickness used in the calculations. A QD $50~\mathrm{nm}$ wide and $15~\mathrm{nm}$ high, representative of the uncapped nanoholes measured by AFM [REF], places the GS transition within the target range but substantially underestimates the ES-GS separation. Reducing the width increases the $l_z\approx=1$ splitting, while increasing the height compensates for the accompanying blueshift of the GS transition. Along this part of the geometry sweep, effective dimensions of $34~\mathrm{nm}$ in width and $31~\mathrm{nm}$ in height bring both quantities near the centers of their target ranges (Figure~\ref{fig:Sims_output}(a,b)).
We constrained the effective QD size using the GS energy and ES-GS splitting measured for Sample~C, whose nominal filling thickness is closest to the $6~\mathrm{nm}$ QW thickness used in the calculations. A QD $50~\mathrm{nm}$ wide and $15~\mathrm{nm}$ high, typical of the uncapped nanoholes measured by AFM \cite{chellu2021highly}, reproduces the GS energy but substantially underestimates the ES-GS separation. Reducing the width increases the $|l_z|\approx1$ splitting but blueshifts the GS, which can be compensated by increasing the height. Following this trend, a width of $34~\mathrm{nm}$ and a height of $31~\mathrm{nm}$ bring both quantities close to the center of their experimental ranges (Fig.~\ref{fig:Sims_output}(a,b)). Unless stated otherwise, this QD size is used in all following calculations.

% InGaSb: Strain and bandedges
%With the QD dimensions fixed at $34~\mathrm{nm}$ in width and $31~\mathrm{nm}$ in height, In incorporation increases the lattice mismatch with the GaSb substrate compresses both the QD and QW. At $12\%$ In, hydrostatic strain reaches $-1.3\%$, while biaxial strain changes sign between the tip and the opening, reaching a minimum of $-3.3\%$ (Figure~\ref{fig:Sims_input}(b,d)). This spatial variation in biaxial strain changes the relative positions of the HH and LH edges along the QD axis. As Figure~\ref{fig:Sims_input}(e,f) shows, the HH edge now rises from the tip toward the opening, while the LH edge falls. The resulting minimum of the hole confinement potential lies on the HH branch in the QW region, drawing the hole GS beyond the nominal QD boundary, while the electron GS remains localized inside the QD (Figure~\ref{fig:Sims_output}(e,f)).

% InGaSb: Energies and spinor nature

Figure~\ref{fig:Sims_output}(c,d) shows how the In content affects the electron and hole energy levels, and Fig.~\ref{fig:Sims_output}(e,f) the GS probability densities at $0\%$ and $12\%$ In. As In narrows the (In)GaSb bandgap, the electron and hole levels approach each other, producing a GS redshift of $4.4~\mathrm{meV}$ per percentage point of In (inset of Fig.~\ref{fig:Sims_output}(d)), i.e. $52.8~\mathrm{meV}$ at $12\%$ In, in excellent agreement with the measured composition dependence of $4.4~\mathrm{meV}/\%$ \cite{peil2026predictive}. The electron excited states comprise a lateral $p$-like pair with dominant $|l_z|\approx1$ character and a state with $l_z\approx0$, as described above. All of them shift nearly in parallel with the GS as In increases, but the splitting between the two $p$-like states grows (Fig.~\ref{fig:Sims_output}(c)). This is consistent with a stronger strain-induced piezoelectric potential, whose octopolar component distinguishes the $[110]$ and $[1\bar{1}0]$ directions and splits the two lateral $p$ orbitals \cite{schliwa2007impact}. At $0\%$ In, several hole states have appreciable LH weight. As In increases, the HH and LH edges separate in the top region of the QD, and these states become predominantly HH-like and form more distinct shell-like groups (Fig.~\ref{fig:Sims_output}(d)). Over the same composition range, the two lowest hole levels approach each other, and the GS hole density evolves from a compact distribution inside the QD to a two-lobed, $p$-like profile in the top region, while the electron density stays compact (Fig.~\ref{fig:Sims_output}(e,f)).

\subsection{Magnetic field calculation}
\label{ssec:results_mag_calculations}

\begin{figure}%[h!]
    \centering
    \includegraphics[width=0.99\linewidth]{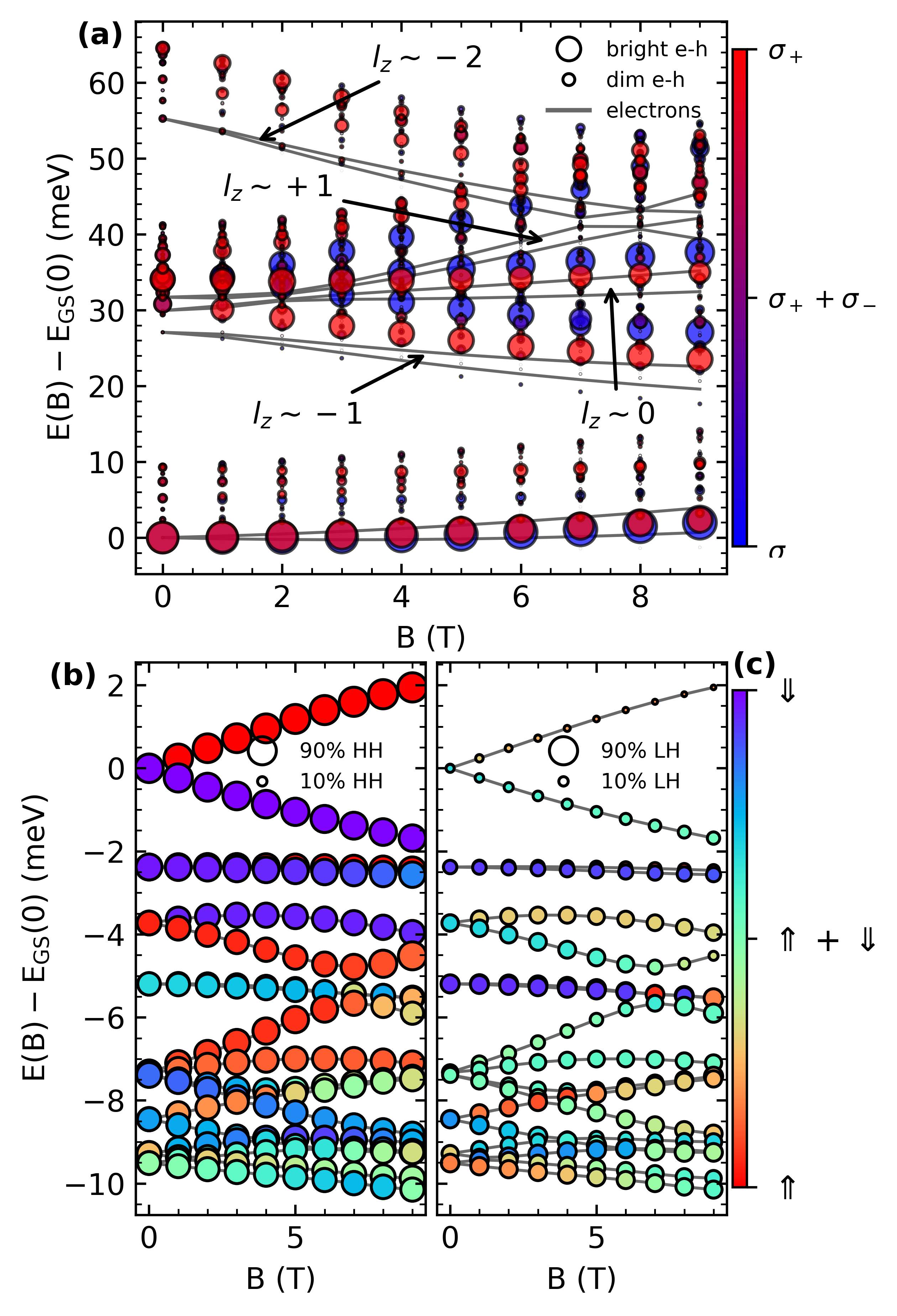}
    \caption{
    Magnetic-field dependence of optical transitions and hole levels calculated using 10 electron states and 18 hole states for an In-free GaSb QD ($34~\mathrm{nm}$ wide, $31~\mathrm{nm}$ high) coupled to a $6~\mathrm{nm}$ GaSb QW, with $\mathbf{B}$ applied along the growth direction. (a) Energies of the electron-hole transitions relative to the zero-field GS transition. Marker size is proportional to the calculated intensity, colour indicates the circular polarization of the emission, and grey lines follow the dispersion of the electron states. (b,c) Energies of the hole states relative to the zero-field hole GS. Marker size gives the HH weight in (b) and the LH weight in (c). Colour indicates the relative weights of the two spin orientations within each component.
    }
    \label{fig:B_sweeps}
\end{figure}

Figure~\ref{fig:B_sweeps} shows the calculated magnetic-field dependence of the electron and hole states and of the resulting optical transitions of the GaSb QD. Panel (a) displays the transition energies together with the dispersions of the lowest 10 electron states (grey lines), which span about $60~\mathrm{meV}$, while panels (b,c) show the 18 lowest hole states, which lie within only about $10~\mathrm{meV}$. Because of this much denser and weaker-dispersing hole spectrum, the field dependence of the transitions is dominated by the electrons, and each transition follows the dispersion of its electron orbital in Fig.~\ref{fig:B_sweeps}(a).

The bright GS transition, which involves the $l_z\approx0$ electron state, shifts upward by only about $2~\mathrm{meV}$ between $0$ and $9~\mathrm{T}$. A much larger field dependence is found for the excited transitions. The two $p$-like electron states move in opposite directions: the $l_z\approx-1$ state shifts downward and the $l_z\approx+1$ state upward, reaching a separation of approximately $25~\mathrm{meV}$ at $9~\mathrm{T}$. The excited $l_z\approx0$ state instead shows a weaker, predominantly diamagnetic shift, while a higher-lying $l_z\approx-2$ state moves down towards the rising $p$-like branch. A Zeeman splitting is also visible within each orbital branch.

Figure~\ref{fig:B_sweeps}(b,c) resolves the hole levels into their HH and LH components. The hole GS doublet remains predominantly HH-like as it splits with field. The excited levels are closely spaced and exhibit both crossings and pronounced anticrossings, with a redistribution of spinor weight between branches that reflects the HH-LH mixing \cite{luttinger1956quantum, chuang1991efficient}. %Despite this dense hole spectrum, a dominant optical transition can be identified for each of the lowest electron states in Fig.~\ref{fig:B_sweeps}(a). The optical dipole matrix element depends on both the electron-hole envelope overlap and the optically allowed spinor components, so energetically close hole states can have very different oscillator strengths. At higher energies, particularly on the rising $l_z\approx+1$ branch, several transitions carry comparable oscillator strength and a single dominant transition is less clearly resolved.

Although many hole states are accessible, the calculated intensity is not distributed over a large number of weak transitions. For the GS and for each of the $p$-like electron states in Fig.~\ref{fig:B_sweeps}(a), a single dominant transition can be identified, which appears as a Zeeman-split doublet at finite field. This follows from the optical dipole matrix element, which depends on both the electron-hole envelope overlap and the optically allowed spinor components, so that energetically close hole states can have very different intensities. At higher energies, particularly on the rising $l_z\approx+1$ branch, several transitions have comparable intensities and the dominant transitions cannot be clearly distinguished.

With increasing In content (not shown), the electron shell structure is preserved, while the hole states acquire predominantly HH character and form more distinct shell-like groups (Fig.~\ref{fig:Sims_output}(c,d)). Since LH admixture mediates the mixing between states of opposite HH spin orientation, this mixing weakens at higher In content, particularly for states in different shells, which are also further apart in energy. The magnetic-field dispersions of the electron and hole states, and hence of the transitions, remain essentially the same, but the hole states show lower LH components.
% With indium

% Overlay to experiment
% Figure~\ref{fig:experiment_theory}(a-c) compares the measured magneto-PL spectra with the calculated optical transitions after applying constant energy offsets of $-8.9$, $-9.3$, and $-11.2~\mathrm{meV}$ for Samples A-C, respectively, to align the GS transitions at zero field. The calculations reproduce the principal magnetic dispersions without further adjustment. Comparison with the fitted transition energies [Figure~\ref{fig:experiment_theory}(d-f)] supports an assignment of the weakly dispersing excited-state component to transitions involving $l_z\approx0$ electron states and of the two more strongly dispersing components to the $l_z\approx\pm1$ electron orbital families.
\section{Discussion} \label{sec:discussion}

\begin{figure*}%[h!]
    \centering
    \includegraphics[width=0.99\linewidth]{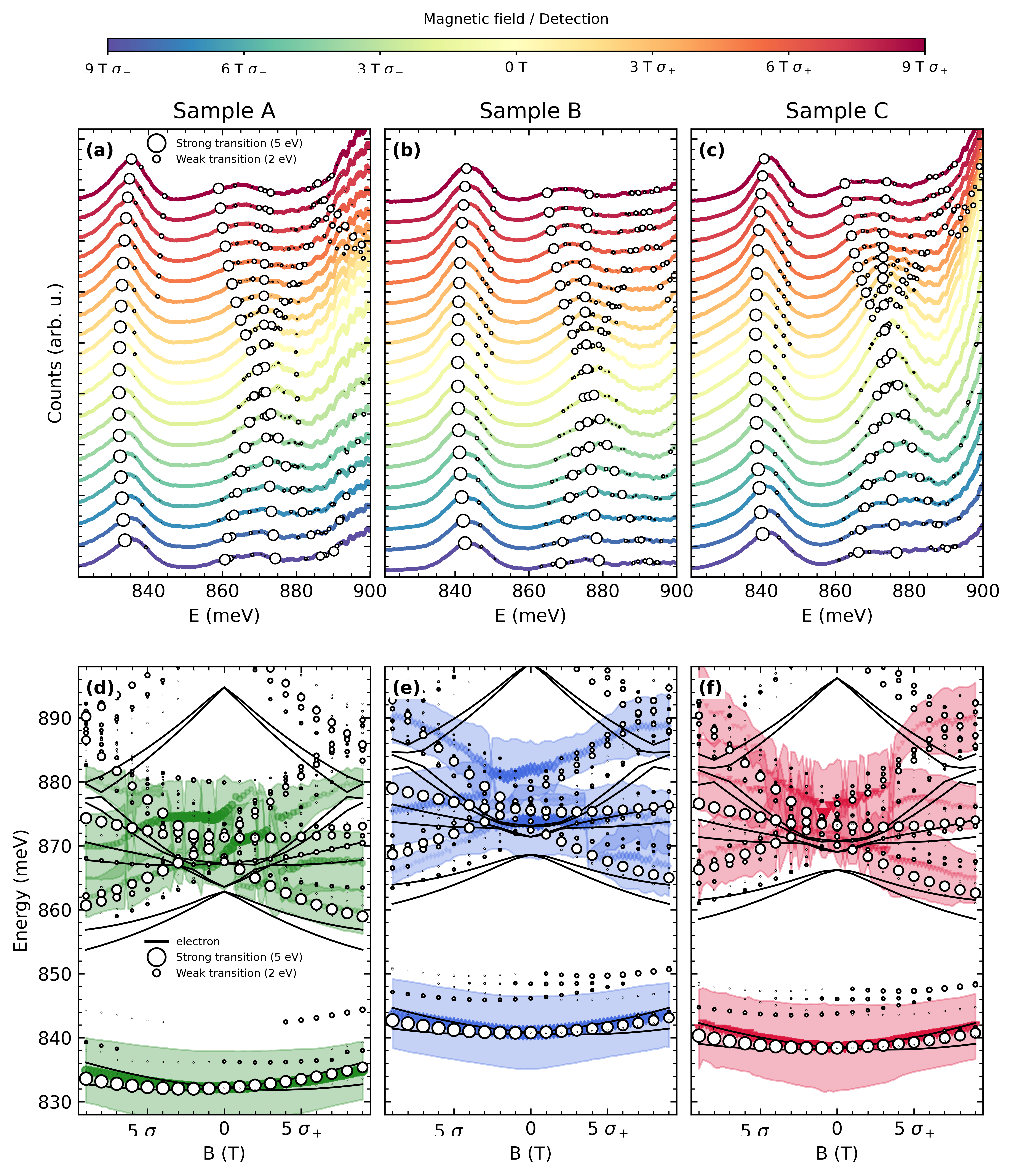}
    \caption{ Comparison of magneto-photoluminescence (PL) experiments and simulations for Samples~A--C (left to right). (a--c) PL spectra at selected magnetic fields, vertically offset from $9~\mathrm{T}$ in $\sigma^{+}$ detection (top) to $9~\mathrm{T}$ in $\sigma^{-}$ detection (bottom). (d--f) Shaded bands trace the fitted GS and three ES components, with a thickness proportional to the component amplitude relative to the GS. Calculated transitions (symbols) are shown for the corresponding field and polarization, and solid lines show the calculated electron-state dispersions. Circles mark stronger and weaker transitions, respectively, with symbol size proportional to the calculated transition intensities. Only transitions satisfying $I_{\sigma^+}+I_{\sigma^-}>1~\mathrm{eV}$ are displayed.
    }
    \label{fig:experiment_theory}
\end{figure*}

% Magnetocomparision intro
We now compare the experimental and numerical results of Sections~\ref{sec:results_magnetopl} and~\ref{sec:results_calculations}. Figure~\ref{fig:experiment_theory} shows the measured magneto-PL spectra of Samples~A--C together with the optical transitions calculated for their respective filling thicknesses and compositions. In the simulations, the QD dimensions were determined from the zero-field GS energy and ES-GS separation of Sample C and were held fixed for all three samples. QW thicknesses are 4.9 nm, 5.7 nm and 6.0 nm for Samples A--C, respectively. Rigid shifts of $-8.9$, $-9.3$, and $-11.2~\mathrm{meV}$ for Samples A--C, respectively, align the calculated GS transitions with the zero-field emission. In panels (a--c), the curves are the QD emission spectra recorded at selected fields in the two circular polarizations. The symbols mark the calculated electron-hole transition energies for the corresponding sample and polarization, with symbol size proportional to the calculated intensity. Each spectrum is fitted with a GS and three ES components of common linewidth; in Fig.~\ref{fig:experiment_theory}(d--f), the thickness of each band is given by $\mathrm{FWHM}\times A_i/A_{\mathrm{GS}}$, where $A_i/A_{\mathrm{GS}}$ is the component amplitude relative to the GS at the same field. The centres of the shaded bands trace the fitted energies of the GS and three ES components. Each band has a displayed thickness of $\mathrm{FWHM}\times A_i/A_{\mathrm{GS}}$, where the four components share a linewidth and the amplitude ratio is evaluated at each field, as explained in Section \ref{sec:results_magnetopl}. The symbols again denote calculated optical transitions, while the solid lines trace the electron-state energies used to identify the orbital character of each branch.

% Magnetocomparision part 2
Across Samples~A--C, the fitted ES energies follow the same ordering as the calculated optical branches: the descending and ascending components track transitions involving the $l_z\approx-1$ and $+1$ electron states, respectively, while the weakly dispersing component tracks the $l_z\approx0$ excitation (Figure~\ref{fig:experiment_theory}(d-f)). The agreement extends to the magnitude of the splitting: at $9~\mathrm{T}$, the outer fitted components are separated by approximately $25~\mathrm{meV}$, comparable to the calculated separation of the two lateral $p$ orbitals. This comparison supports assigning each fitted component to a family of transitions sharing the same electron shell. The correspondence is least precise for the ascending branch, where several calculated transitions carry comparable intensities, as mentioned above, and the measured QD emission overlaps the low-energy tail of the P3 peak. 

% Effective confinement geometry
The same QD dimensions that account for the zero-field spectrum of Sample C also reproduce the principal ES dispersions across all three samples. These dimensions (34 nm wide, 31 nm high)  differ appreciably from those measured by AFM on uncapped nanoholes (50 nm wide, 12 nm depth), whose profiles may change upon filling and capping. Cross-sectional investigations of GaAs/AlGaAs QDs have suggested steeper buried side facets than those deduced from earlier AFM nanohole characterization \cite{nemcsics2011cross}, while three-dimensional reconstructions have revealed asymmetric morphologies and chemically nonuniform, Al-rich sidewalls \cite{zhang2024unveiling}. Similar Al enrichment has been observed at GaSb/AlGaSb QD interfaces \cite{masson2026engineering}, and calculations have shown that interface smoothing can appreciably modify the confined optical transitions \cite{mlinar2007theoretical, peil2026predictive}. Although the inverted-conical geometry adopted here does not explicitly include these structural details, its effective dimensions can still capture their combined influence on the electronic confinement. 

% Non-resonant excitation
The relative intensities and fitted energies in Fig.~\ref{fig:experiment_theory} also depend on how the emitting states are populated. Under strong non-resonant excitation, several multicarrier configurations can contribute to the same inhomogeneously broadened QD emission band, and their occupation depends on the carriers supplied by the surrounding QW. Moreover, disorder-mediated relaxation can modify the carrier spin populations before capture \cite{barbieri2026magneto}, and magnetic-field-dependent transport can alter the QD feeding \cite{larsson2006magnetic}. The resulting redistribution among overlapping optical transitions can change the amplitudes and energies of the fitted components, including the GS. These changes add to the intrinsic magnetic shifts and to the changes in optical selection rules. Although the main orbital dispersions can be identified, such population effects prevent a reliable determination of the excitonic diamagnetic coefficients or $g$-factors from the present spectra.

% Hole manifold
The closely spaced hole states found in our calculations (see Fig.\ref{fig:Sims_output}(b)), whose dense spectrum is also supported by quasi-resonant spectroscopy of individual InGaSb/AlGaSb QDs \cite{hakkarainen2026telecom}, provide a pathway for the captured excitonic complex to relax before recombination. Owing to their substantial HH-LH mixing, this relaxation may change the spin character of the emitting complex, introducing a potential limitation for hole-spin applications \cite{warburton2013single}. Such a mechanism has already been implicated in GaAs QDs, where phonon-assisted transitions involving mixed excited states contribute to the degradation of photon-pair entanglement \cite{lehner2023beyond}. The stronger HH character and larger energy separations of the excited hole states obtained here with In incorporation may therefore favour spin preservation by reducing both spin mixing and thermal access to excited states.

% Effects of Indium on hole localization
Increasing the In content also leads to a loss of hole localization in the QD. The evolving HH band-edge profile (Fig.~\ref{fig:Sims_input}(f)) favours the localization of the holes in the surrounding QW, which reduces the electron--hole overlap and weakens the lateral confinement of the hole. In our calculations, this is accompanied by the development of a two-lobed GS probability density in the top region of the QD (Fig.~\ref{fig:Sims_output}(f)). Although the details of this feature may depend on the idealized geometry, the band-edge evolution supports the underlying trend toward hole spillout into the QW.

% g-factor manipulation

The displacement of the hole towards the QW also suggests a route to tuning its magnetic response. Confinement can quench orbital angular momentum and modify band mixing, producing carrier $g$-factors whose magnitude, sign and anisotropy differ substantially from their bulk values \cite{pryor2006lande, van2016anisotropy}. In GaAs/AlGaAs LDE QDs, this sensitivity has been used to predictively engineer electron and hole magnetic properties through structural design, including tuning the in-plane electron $g$-factor through zero \cite{schimpf2025optical}. The strong spin-orbit coupling and large bulk $g$-factors of antimonides suggest that similar control may be possible in (In)GaSb/AlGaSb QDs. Electrical tuning of excitonic $g$-factors through changes in hole localization and orbital character has also been demonstrated in InGaAs and InAs/GaAsSb QDs \cite{jovanov2011observation, llorens2019topology}, and could be combined with a graded In profile, so that the hole samples a different local composition and strain.

\section{Conclusions}
\label{sec:conclusion}

We have investigated the electronic structure of (In)GaSb/AlGaSb QDs, a promising platform for single-photon emission at telecom wavelengths, by combining magneto-photoluminescence with k·p calculations. The ground- and excited-state energies constrain an effective QD geometry (34 nm wide, 31 nm high), whose calculated magnetic-field dependence reproduces the main experimental dispersions in three samples. This agreement allows the main excited-state features to be assigned to distinct electron orbital families, while the transitions are accompanied by a dense manifold of closely spaced, mixed hole states.

In alloying modifies this electronic structure beyond shifting the emission wavelength. The increased strain-induced HH-LH separation favors predominantly HH-like hole states, but the accompanying changes in the confinement potential progressively shift the hole localization towards the surrounding QW. At higher In contents, this may compromise the lateral hole confinement, an important consideration for the design of telecom quantum emitters. At the same time, this sensitivity of the hole wavefunction suggests routes to tune the magneto-optical properties of these QDs, which is relevant for spin-photon interfaces. Quasi-resonant, charge-resolved magneto-optical spectroscopy of individual QDs would help link the electronic structure identified here to specific excitonic configurations and their magnetic properties.

\section*{Acknowledgements}

This publication is part of the R\&D\&I project PID2022-136626OB-C33, funded by MICIU/AEI/10.13039/501100011033 and by ERDF, EU. We also acknowledge financial support from the European Union's Horizon 2020 research and innovation programme under the Marie Sk\l{}odowska-Curie grant agreement No.~956548 (project Quantimony); the CSIC Interdisciplinary Thematic Platform (PTI+) on Quantum Technologies (PTI-QTEP+); the Research Council of Finland projects QuantSi (Decision No.~323989) and CryoLight (Decision No.~357351) and Flagship Programme PREIN; and Business Finland, through the co-innovation project QuTI (41739/31/2020) and the Rise to Challenge project TeleQuant (1835/31/2025).

\bibliographystyle{unsrt}
\bibliography{sample}

\end{document}